\documentclass[pra,showpacs,preprint,superscriptaddress,amsmath,amssymb]{revtex4-1}

\setcitestyle{round}

\usepackage{graphicx}
\usepackage{dcolumn}

\usepackage[usenames]{color}

\begin{document}
\raggedbottom

\title{Calculation of molecular vibrational spectra on a quantum annealer}
\author{Alexander Teplukhin}
\author{Brian K. Kendrick}
\email{Correspondence should be addressed to BKK (bkendric@lanl.gov).}
\affiliation{Theoretical Division (T-1, MS B221), Los Alamos National Laboratory, Los Alamos, New Mexico 87545, USA}
\author{Dmitri Babikov}
\affiliation{Department of Chemistry, Marquette University, Milwaukee, Wisconsin 53021, USA}
\vskip 10pt
\begin{abstract}
Quantum computers are ideal for solving chemistry problems due to their polynomial scaling 
with system size in contrast to classical computers which scale exponentially. Until now 
molecular energy calculations using quantum computing hardware have been limited to quantum 
simulators. In this paper, a new methodology is presented to calculate the vibrational spectrum 
of a molecule on a quantum annealer. The key idea of the method is a mapping of the ground 
state variational problem onto an Ising or quadratic unconstrained binary optimization (QUBO) 
problem by expressing the expansion coefficients using spins or qubits. The algorithm is general 
and represents a new revolutionary approach for solving the real symmetric eigenvalue problem on 
a quantum annealer. The method is applied to two chemically important molecules: O$_2$ (oxygen) 
and O$_3$ (ozone). The lowest two vibrational states of these molecules are computed using both 
a hardware quantum annealer and a software based classical annealer.
\end{abstract}

\maketitle

\section*{Introduction}

Quantum computers are seen by many as a future alternative to classical computers. 
Although quantum supremacy has not yet been achieved, the field is advancing quite rapidly. 
There are three major types of quantum computing devices \cite{qcreview}: quantum annealer 
\cite{dwave}, quantum simulator \cite{stancil, aguzik, qsimreview}, and a universal quantum 
computer. The first type is an example of adiabatic quantum computing \cite{aqc} and is used 
to solve optimization problems, which at first glance appears to be quite restrictive. 
The second type is based on quantum gates which appears to have a wider applicability and 
therefore may be able to simulate a larger variety of problems. However, adiabatic and gate 
based quantum computing were proven to be formally equivalent \cite{equiv}. Thus, the practical 
application space is most likely limited by the hardware realization and not necessarily 
by the type of approach. The third type will be able to solve any problem but it does not yet 
exist. Building a universal quantum computer is a very challenging task and its realization 
may be decades away.

Coming from the physical chemistry community, we asked ourselves if it would be possible to 
program an important fundamental problem on a quantum annealer such as the commercially 
available D-Wave machine \cite{dw2000Q}. Typically, people who work with such devices go in 
the opposite direction: knowing hardware capabilities they come up with a suitable optimization 
problem. As a fundamental problem we chose to calculate the vibrational ground state and 
possibly excited states of a molecule. This problem is very important in chemistry, for 
example: 
H$_n^+$ ions \cite{hions1, hions2, hions3, hions4}, 
CH$_5^+$ and isotopologues \cite{ch5-1, ch5-2}, 
H$_3$O$^+$, H$_5$O$_2^+$ and deuterated analogues \cite{h3o, h5o2}, 
hydrogen clusters \cite{h2clust1, h2clust2, h2clust3}, 
their isotopologues \cite{d2clust1, d2clust2}, 
hydrogen bonded systems \cite{hbs1} 
and Lennard-Jones clusters \cite{lj1, lj2}.
The common method to study these molecular systems is a Monte-Carlo (MC) method in its various 
flavors: variational MC, time-dependent variational MC, diffusion MC, and path integral MC.

A very similar ground state problem was addressed recently \cite{vqe1,vqe2,vqe3}, where the 
authors designed an algorithm to calculate the electronic ground state on a quantum simulator. 
The result of their work is a Variational Quantum Eigensolver (VQE), where an expectation 
value of each term in the electronic Hamiltonian is evaluated on a trial wave function using 
a quantum simulator and the resultant total energy serves as a guide for generating the next 
trial wave function. The method is hybrid, because an optimization step, namely the trial 
generation, is performed on a classical computer. 

In contrast to the VQE algorithm which is based on a quantum simulator, the new algorithm 
presented in this work is based on a quantum annealer and solves any real symmetric eigenvalue 
problem. To our knowledge, this is the first quantum annealer based eigenvalue solver and
will be referred to below as the Quantum Annealer Eigensolver (QAE). As discussed in more 
detail below, our QAE algorithm is also hybrid since the variational eigenvalue problem 
is solved via a sequence of many quantum annealer optimizations performed with varying weights 
on the constraint equations (i.e., Lagrange multipliers). The scanning and optimization of the 
weights is done on a classical computer. Mapping the eigenvalue problem to a quantum annealer
hardware is non-trivial, because the annealer solves a minimization problem defined by an Ising 
functional of the form 
$H(s) = \sum_i h_{i} s_i + \sum_{i<j} J_{ij} s_i s_j$, 
where the spin variables $s_i$ accept values \{-1,1\}. Alternatively, the functional can be 
converted to quadratic unconstrained binary optimization (QUBO) form using variables 
$x_i \in \{0,1\}$, called qubits, 
giving $H(x) = \sum_i Q_{ii} x_i + \sum_{i<j} Q_{ij} x_i x_j$ \cite{dwProg}. The problem is how 
to write down a ground state or eigenvalue problem in QUBO form and explicitly construct the 
matrix $\mathbf{Q}$.

The outline of the paper is as follows: First, we present our solution to this problem, 
including the extension to the excited state calculations and multiple dimensions. Second, we apply 
our algorithm to two chemically important species, O$_2$ (oxygen) and O$_3$ (ozone). Third, the 
introduction of weighted constraints is presented following a technique used to overcome the 
connectivity issue in the quantum annealer hardware (i.e., D-Wave machine). Noise is also modeled 
in the algorithm which is shown to reproduce the results from the D-Wave machine. In the final 
discussion section, we consider possible improvements of the algorithm and sources of error.

\section*{Results}

\subsection*{Mapping of ground state problem to QUBO problem}

The method is inspired by the variational principle. Suppose we are interested in a ground 
state of a one-dimensional system and its wave function $\Psi$ is expanded using an orthonormal 
basis $\varphi_\alpha$ and unknown expansion coefficients 
$a_\alpha$: $\Psi = \sum_{\alpha=1}^B a_\alpha \varphi_\alpha$. 
Then, the ground state energy can be expressed as a double sum over the Hamiltonian 
matrix elements 
$E = \langle\Psi\vert \hat H \vert\Psi\rangle = 
\sum_{\alpha,\beta}^{B,B} a_\alpha a_\beta
\langle\varphi_\alpha\vert \hat H \vert\varphi_\beta\rangle = 
\sum_{\alpha,\beta}^{B,B} a_\alpha a_\beta H_{\alpha\beta}$
It is easy to see that the functional form for the energy $E$ is similar to QUBO form $H(s)$, 
except that the coefficients $a_\alpha$ are continuous and $a_\alpha \in [-1;1]$ (since from 
the normalization condition $\langle\Psi\vert\Psi\rangle = 1$ we know 
$\sum_{\alpha=1}^B\,a^2_\alpha = 1$). In contrast, the QUBO variables $x_i$ are discrete.

The key idea in mapping the eigenvalue problem with Hamiltonian matrix $\mathbf{H}$ to the 
QUBO optimization problem with matrix $\mathbf{Q}$ is to express each expansion coefficient 
$a_\alpha$ using $K$ qubits $q_k^\alpha$. This approximation can be done in multiple ways. 
The approach we followed in this work is a fixed-point representation that is used to 
represent real numbers in a computer. Since the magnitude of the coefficients $a_\alpha$ never 
exceeds unity, only the fractional part of the coefficient has to be stored. The last qubit 
$q_K^\alpha$ stores the sign of $a_\alpha$. The complete expression for the coefficient is 
$a_\alpha = \sum_{k=1}^{K-1}2^{k-K}q_k^\alpha - q_K^\alpha \in [-1;1)$. Now, the functional 
$E$ can be expressed explicitly in terms of the qubits $q_k^\alpha$. The powers of two are 
combined with the matrix elements $H_{\alpha\beta}$ giving the matrix elements $Q_{ij}$. 
The qubits $q_k^\alpha$ are mapped to the qubits $x_i$ via the relation $i = K(\alpha-1) + k$, 
where $i \in [1, B \times K]$.
Since in the QUBO (or Ising) model the ordering of qubits within the pair does not matter 
(i.e., the interaction between $i$ and $j$ is the same as $j$ and $i$), the summation is 
restricted to $i < j$ and the non-diagonal elements $Q_{ij}$ are multiplied by two.

Unfortunately, the minimum of the functional $E$ is a trivial solution $\Psi = 0$, which is 
due to the lack of the normalization constraint $\|\Psi\| = 1$. The workaround is to add 
that constraint right into the functional with a strength $\lambda$, giving 
$I = \langle\Psi\vert \hat H \vert\Psi\rangle + \lambda(1-\langle\Psi\vert\Psi\rangle)^2$. 
Essentially, the parameter $\lambda$ penalizes any deviation of the norm from unity and it
helps to guide the optimization away from the trivial solution. One can think of $\lambda$ 
as a Lagrange multiplier and the functionals $E$ and $I$ as objective functions. The problem 
with the functional $I$ is that it is no more a QUBO functional, rather it is biquadratic 
in $x$. The trick is to lower the power of the constraint, giving 
$G = \langle\Psi\vert \hat H \vert\Psi\rangle + \lambda(1-\langle\Psi\vert\Psi\rangle)$.
Dropping the constant shift $\lambda$, which has no effect on the optimization, one obtains 
the final expression for the functional from used in present study: 
$F = \langle\Psi\vert \hat H \vert\Psi\rangle - \lambda\langle\Psi\vert\Psi\rangle$.
The main consequence of the decreased power is that the normalization condition is broken 
\textit{per se} (but this can be fixed by renormalizing the final solution). However, the 
primary role of the penalty is to avoid the trivial solution and the functional $F$ serves 
that purpose. Another issue with the functional $F$ is that it encourages a nonphysical 
norm $\|\Psi\| > 1$. This limits the number of techniques to find a good parameter $\lambda$. 
Ideally, $\lambda$ should be large enough to kick the optimization away from the trivial 
solution minimum but yet small enough to stay away from the large norm limit. To find an 
optimal value for $\lambda$, we scan in $\lambda$ and pick the solution with the lowest 
energy $E=\langle\Psi\vert \hat H \vert\Psi\rangle$ (where here $\Psi$ has been
renormalized: $\langle\Psi\vert\Psi\rangle=1$).

\subsection*{Calculation of excited states}

The QAE algorithm described above can be easily applied to the calculation of excited states 
by modifying the initial Hamiltonian $\mathbf{H}$. Specifically, for the first excited state, 
an outer product matrix of the previously computed ground state wave function is added: 
$\mathbf{H'} = \mathbf{H} + S_0 \vert\Psi_0\rangle\langle\Psi_0\vert$. 
The parameter $S_0$ is an arbitrary user specified energy shift to move the ground state 
higher in the spectrum. The only requirement on $S_0$ is that it should be larger than the 
energy of the first excited state, otherwise the algorithm will keep converging to the ground 
state. To compute the $i$-th excited state, similar terms for the states $0,\ldots,(i-1)$ 
should be added to the Hamiltonian. In principle, this iterative procedure allows one to 
compute the whole spectrum of a molecule. Obviously, for a fixed basis size $B$ and qubit 
expansion $K$, the higher states will not be described as accurately as the lower ones.

\subsection*{Multiple dimensions}

The generalization of the QAE method to multiple dimensions is straightforward. For a direct 
product basis, the one-dimensional expansion is replaced with an $n$-dimensional expansion, 
but the way to code each expansion coefficient using $K$ qubits remains the same. For 
example, for a two-dimensional system with the same number of basis functions $B$ for each 
dimension, the expansion is 
$\Psi = \sum_{\alpha,\beta}^{B,B} a_{\alpha\beta} \varphi_\alpha \theta_\beta$, 
where $\varphi_\alpha$ and $\theta_\beta$ are the basis functions in each dimension. The 
qubits $q_k^{\alpha\beta}$ are now mapped to the variables $x_i$ as follows: 
$i = KB(\alpha-1) + K(\beta-1)+ k$, where $i \in [1, B^2 \times K]$.

The QAE method can also be applied to a non-direct product basis. For example, one can 
implement a Sequential Diagonalization Truncation (SDT) \cite{sdt1, sdt2, sdt3} which 
drastically reduces the size of the Hamiltonian matrix and ultimately results in a much 
smaller total number of qubits than in direct product treatment. We use the SDT approach 
for ozone and further details of applying the SDT method to that molecule can be found 
elsewhere\cite{ozone1}.

\subsection*{Application to O$_2$ and O$_3$}

\begin{figure}[h]
\includegraphics[scale=0.5]{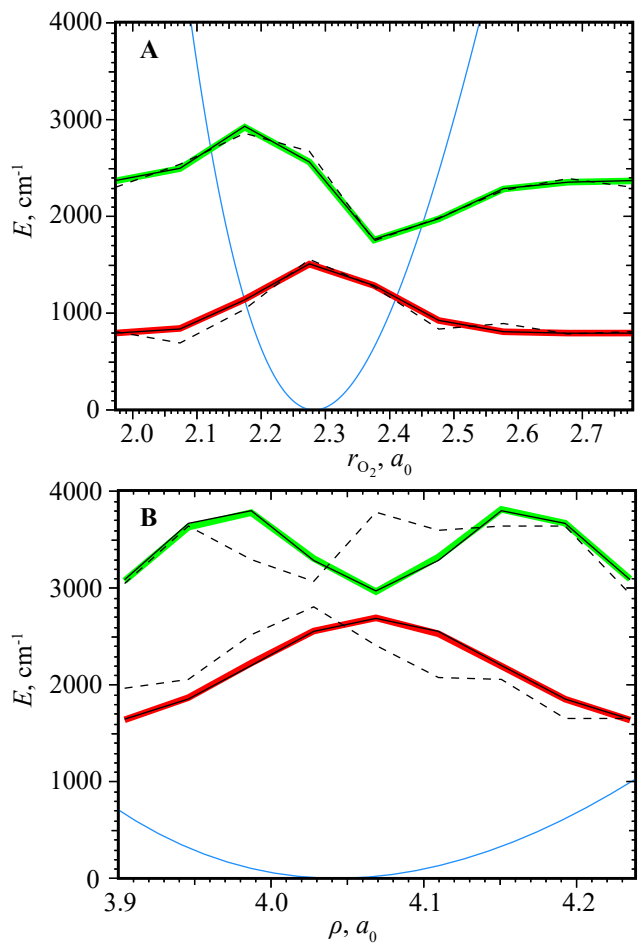}
\caption{
\begin{small} {\textbf{Computed wave functions of the ground and first excited states.} The molecules are (\textbf{A}) O$_2$ and (\textbf{B}) O$_3$, the wave function is squared for ozone. Red and green bold curves are the results of a classical numerical eigensolver (LAPACK). Solid thin black curves were obtained using QAE with a software classical QUBO solver (they lie on top of the LAPACK curves). Dashed black curves are the results obtained using QAE with a hardware quantum annealer (D-Wave machine). The blue curve is (A) oxygen potential $V(r_{{\rm O}_2})$ and (B) minimum energy path for ozone.}
\end{small}
}
\label{fig01}
\end{figure}

We applied our algorithm to the calculation of the ground and first excited states of 
the oxygen and ozone molecules. For both, we used an accurate potential energy surface of 
ozone \cite{ozonepes}. The one-dimensional O$_2$ potential $V(r_{{\rm O}_2})$ was generated 
from the full three-dimensional ozone potential by moving one of the oxygen atoms far away 
from the other two (i.e., $R_{{\rm O}-{\rm O}_2} = 60$ $a_0$). The two lowest wave functions 
for both molecules are shown in Fig. \ref{fig01}. The software based classical QUBO solver 
reproduces the wave functions computed with a standard classical numerical eigensolver 
(LAPACK \cite{lapack}), whereas the output of the hardware quantum annealer (D-Wave machine) 
is less accurate. The sharp edges (low resolution) of the wave functions is due to the small 
size of the basis set. For oxygen, a Fourier basis ($e^{i\,m\,r_{O_2}}$) was chosen small 
enough to give the known ground state energy of 791.64 cm$^{-1}$ within an error of 0.01 
cm$^{-1}$ using a standard classical numerical eigensolver (LAPACK). The required number of 
periods in the Fourier series is $m_{max} = 4$ which translates to the basis size 
$B = 2\,m_{\max}+1 = 9$. For ozone, an accurate calculation would require significant 
computational resources, so we decided to use an SDT basis set truncated at quite low energy 
$E_{cut} = 2000$ cm$^{-1}$, which gives $B = 12$ basis functions. This basis is sufficient 
to describe the ground state at 1451 cm$^{-1}$ with an error of 120 cm$^{-1}$ using LAPACK, 
but is too small for excited state calculations (the computed excited state energy is 700 
cm$^{-1}$ larger than the true value of 2147 cm$^{-1}$). Nevertheless, in this work we are 
primarily interested in benchmarking the new QAE method against a standard classical numerical 
eigensolver (LAPACK) and not in the absolute accuracy of the solutions. We refer the reader 
interested in accurate classical calculations to the relevant literature \cite{ozone1, ozone2}. 
For the three-dimensional ozone system, we plot the probability density (the wave function squared)
as a function of the symmetric-stretch coordinate $\rho$ in Fig. 1b \cite{ozonecoords1, 
ozonecoords2}. The SDT basis functions span the other two internal degrees of freedom (not 
plotted) at each value of $\rho$ and are computed classically \cite{ozone1}. The number of 
qubits $K$ per expansion coefficient (or basis function) is 7 for oxygen and 5 for ozone 
which is the maximum possible number which fits within the 64 logical fully-connected qubits 
on the hardware quantum annealer (D-Wave machine). Namely, $K \cdot B = 7 \cdot 9 = 63$ for 
oxygen and $K \cdot B = 5 \cdot 12 = 60$ for ozone.

Figure \ref{fig02} illustrates the convergence of the energies as a function of the number of 
qubits $K$ per expansion coefficient $a_\alpha$ (i.e., the level of discretization). There 
are several interesting findings to discuss. First, the error decreases exponentially as a 
function of $K$, which is very appealing. Second, the error decreases and reaches a plateau, 
for both solvers and both systems. This shared behavior demonstrates that the QAE algorithm 
itself is universal, but the actual error is solver and system dependent. Third, the quantum 
annealer (D-Wave machine) is much less accurate (by two-three orders of magnitude) than the 
classical QUBO solver. In addition, because the total number of qubits currently available in 
the quantum annealer is rather limited, the corresponding (dashed) curves do not continue to 
higher values of K. Finally, the classical QUBO solver brings the error for O$_2$ down to 0.01 
cm$^{-1}$ which coincidently matches the error of chosen basis size. No more than $K = 8$ 
qubits (a qubyte) per coefficient are needed for oxygen. For ozone, $K = 5$ qubits are required 
by the classical QUBO solver to reach the plateau within an error of less than 3 cm$^{-1}$. 
This is quite accurate, compared to the 120 cm$^{-1}$ error due to the chosen (small) SDT basis. 

Table S1 in Supplementary Materials gives the parameters used in scanning over the normalization 
penalty $\lambda$. The initial $\lambda_{min}$ could be zero or some value that is smaller than the 
expected energy of the state being computed (if it is known). The number of steps $N_\lambda$ 
specifies the number of samples and is a convergence parameter. The last parameter is the 
step size $\Delta\lambda$ which ideally should be the same for all $K$ within a given problem. 
However, for small $K$ we were always getting trivial solutions which is probably due to the 
inaccurate description of the problem. The only way we found to avoid this is to increase the 
step size $\Delta\lambda$ by more than an order of magnitude. We faced the same issue when we 
were simulating the hardware noise. Increasing both the step size and the number of steps allowed 
us to overcome this obstacle.

\subsection*{Algorithm scaling}

It is straightforward to show that the runtime scales as $O(N_\lambda KB^d)$, where $d$ is the 
number of dimensions and $B$ basis functions are used for each dimension. In practice, however, 
the QUBO solver may also have some additional (internal) effects on performance. For example, 
the classical QUBO solver we used in this work is based on a backbone-based method inspired by 
Glover, \textit{et al.} \cite{glover} to partition the problem, which is causing a step-like 
runtime as a function of $K$ (see Figure S1 in Supplementary Materials). In the next paragraph, 
we will show that the actual computational time is in agreement with the theoretical scaling for 
the $d$-dimensional harmonic oscillator problem solved using a cosine basis and the classical 
QUBO solver. The oscillator frequencies were set different according to 
$\omega_i ($cm$^{-1}) = 800 + 200 \cdot (i-1)$, where $i$ is the dimension index. The linear 
scaling with $N_\lambda$ is obvious. 

To verify the linear scaling with $K$, we plot the normalized computational time divided by $K$
as a function of $K$ in Figure \ref{fig03}. As expected, all curves are horizontal and their 
slope does not depend on dimensionality (except for perhaps $d=5$). The rapid increase or steps 
at $K=16$ for 1D, at $K=6$ for 2D and at $K=2$ for 3D are due to partitioning size. They appear 
as soon as the total number of qubits exceeds the sub-QUBO size of 47, which is $3 \times 16$ 
for 1D, $3^2 \times 6$ for 2D and $3^3 \times 2$ for 3D. Once the linear dependence on $K$ has 
been established (see Fig. \ref{fig03}), one can then verify the exponential dependence on 
dimensionality $d$. The logarithm of the normalized computational time is plotted in Figure 
\ref{fig04} as a function of the dimensionality $d$. All of the curves exhibit a linear 
dependence on $d$ which confirms the exponential scaling. Again, the times for all $K$ at $d = 1$ 
and for $K = 4$ at $d = 2$ deviate significantly from the main trend because no partitioning is 
required to compute those. The average slope calculated based on $d=2$ through $4$ is 0.58 
which is close to the theoretically predicted value of $0.48$ (for $B=3$). When $d=5$ is included 
the slope increases to 0.72 which is most likely due to the very large problem size. The total 
number of qubits for $d=5$ is $KB^d = 972$ to 3888 for $K=4$ to 16 and $B=3$ which results in 
a total number of configurations $10^{300}$ to $10^{1000}$. In summary, Figures \ref{fig03} and 
\ref{fig04} confirm the general scaling law $O(N_\lambda KB^d)$ of the algorithm and any 
deviations from it are due to a particular QUBO solver.

We did not perform scaling studies on the hardware quantum annealer (D-Wave machine) because it 
was not practical due to the small number of logical qubits and long runtime. For example, for 
O$_2$ we were able to approach small problems with $B = 9$ and $K = 1$ to 7 and the runtime was 
about $t_{dw} = 2500$ s. This time does not depend on the number of logical qubits $K$, because 
all problems are treated by the hardware as maximum-size problems. In contrast, the classical 
QUBO solver runtime for these problems was about $t_{cl} = 30$ s, which is almost two orders of 
magnitude smaller. The long runtime for the hardware quantum annealer is primarily due to the 
large number of reads (see Fig. S3). In addition, the analysis for the $d$-dimensional 
harmonic oscillator would require an extensive QUBO partitioning, which means another factor of 
ten to hundred increase in $t_{dw}$.

\subsection*{Chaining in quantum annealer}

The only constraint we have discussed so far is a normalization constraint with associated 
penalty $\lambda$. However, there is another constraint and penalty factor worth mentioning 
when running on a quantum annealer (D-Wave machine). Namely, the chain constraint and the 
associated chain penalty. The physical qubits in the hardware do not have an all-to-all 
connectivity which is a requirement of the algorithm. In fact, each qubit has six neighbors 
at most (see Chimera graph) \cite{dwProg}. Fortunately, there is a method to embed a fully 
connected graph on top of the hardware graph. In this approach, a number of qubits are 
organized into so-called chains. Qubits within a chain act like a single logical qubit which 
is connected to all other logical qubits. To program chains in the hardware Chimera graph, 
one adds a set of constraints which have a single strength or chain penalty $c$. As with any 
constraint, the associated penalty or weight $c$ should be neither too small, because then the
chains are broken, or too large, because then the hardware will become insensitive to the 
original problem. The simplest approach to find a good chain penalty is to perform scanning, 
in a similar way to $\lambda$ scanning. Figure \ref{fig05} demonstrates an example of this 
two-dimensional scanning for the ground state of O$_2$ with a Fourier basis of size $B = 7$ 
($m_{max} = 3$) and $K = 3$. As expected, in the region of small $c$ the minimum energy is 
unacceptably large, simply due to broken chains. For large $c$ we see two regions: the region 
of small $\lambda$, which contains trivial solutions only (because the normalization 
constraint is too weak) and the region of large $\lambda$ with reasonable minimum energies. 
The phase transition between these two regions occurs close to the true ground state energy 
791.64 cm$^{-1}$. The result of this two-dimensional scanning shows that a reasonable chain 
penalty lies between 10000 - 20000 cm$^{-1}$. We used $c=$15000 cm$^{-1}$ in all of our 
calculations.

\subsection*{Simulating hardware noise}

\begin{figure}[h]
\includegraphics[scale=0.5]{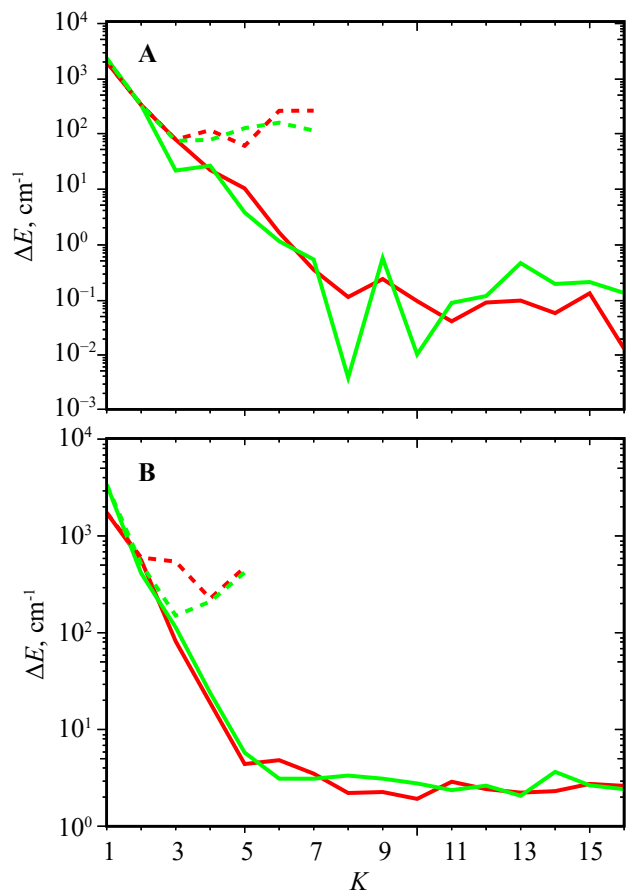}
\caption{
\begin{small} {\textbf{Ground and first excited state energy errors as a function of number of qubits \textit{K} per expansion coefficient \textit{a}$_\alpha$.} The molecules are (\textbf{A}) O$_2$ and (\textbf{B}) O$_3$.  Ground state is red, excited state is green. Results were obtained using QAE with a classical QUBO solver (solid curves) and a quantum annealer (dashed curves). Errors are computed relative to the energies of a classical numerical eigensolver (LAPACK).}
\end{small}
}
\label{fig02}
\end{figure}

\begin{figure}[h]
\includegraphics[scale=0.5]{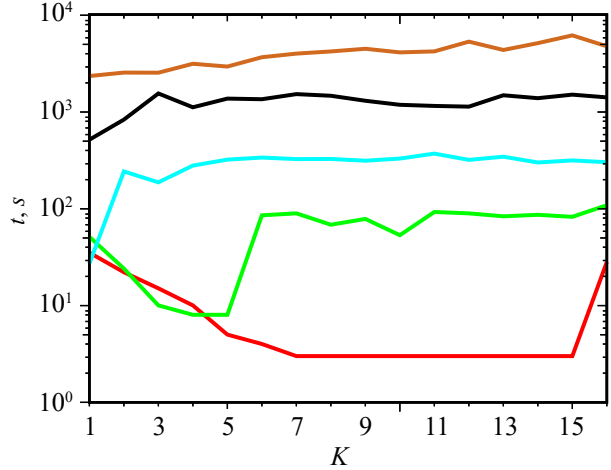}
\caption{
\begin{small} {\textbf{Normalized computational time for the \textit{d}-dimensional harmonic oscillator.} The time is plotted as a function of $K$ and was obtained using QAE with a classical QUBO solver. Results are presented for 1D (red), 2D (green), 3D (blue), 4D (black) and 5D (brown). Number of cosine basis function is $B = 3$ ($m_{max} = 2$). No slope in all curves demonstrates linear scaling of the algorithm with $K$.}
\end{small}
}
\label{fig03}
\end{figure}

\begin{figure}[h]
\includegraphics[scale=0.5]{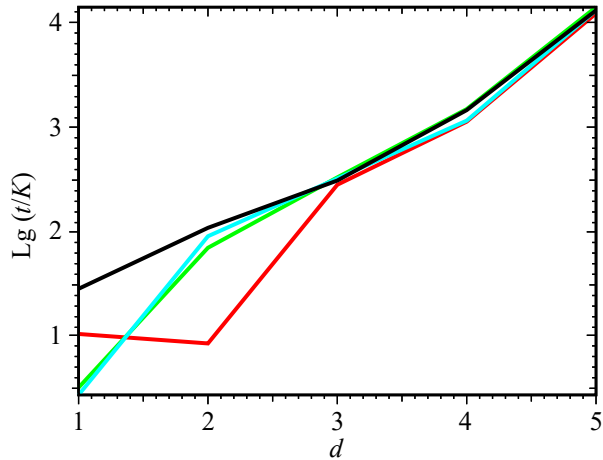}
\caption{
\begin{small} {\textbf{Logarithm of the normalized computational time as a function of the number of dimensions.} The time was measured for the $d$-dimensional harmonic oscillator and was obtained using QAE with a classical QUBO solver. Results are presented for the number of qubits per coefficient $K = 4$ (red), $K = 8$ (green), $K = 12$ (blue) and $K = 16$ (black). Number of cosine basis functions is $B = 3$ ($m_{max} = 2$). The linear dependence confirms exponential scaling in $d$.}
\end{small}
}
\label{fig04}
\end{figure}

\begin{figure}[h]
\includegraphics[scale=1.0]{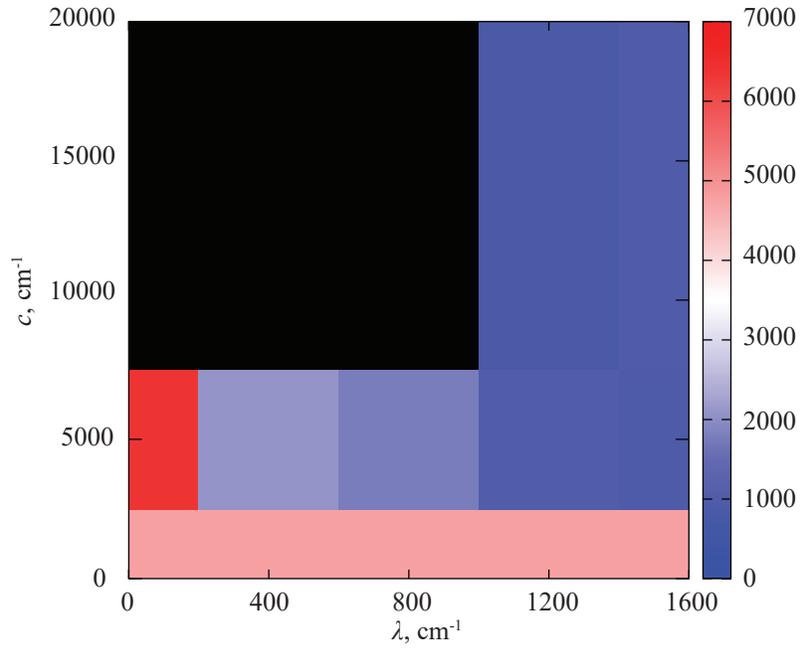}
\caption{
\begin{small} {\textbf{Scanning in the normalization $\lambda$ and chain $c$ penalties.} The calculation was performed on the hardware quantum annealer (D-Wave machine) to find the minimum energy (see color scale in cm$^{-1}$). The three major regions of interest are: trivial solutions (top-left black), broken chains (bottom red-blue) and the optimal region where {\it both} the normalization and chain constraints are satisfied and the minimum energy solution is obtained (top-right blue).}
\end{small}
}
\label{fig05}
\end{figure}

\begin{figure}[h]
\includegraphics[scale=0.8]{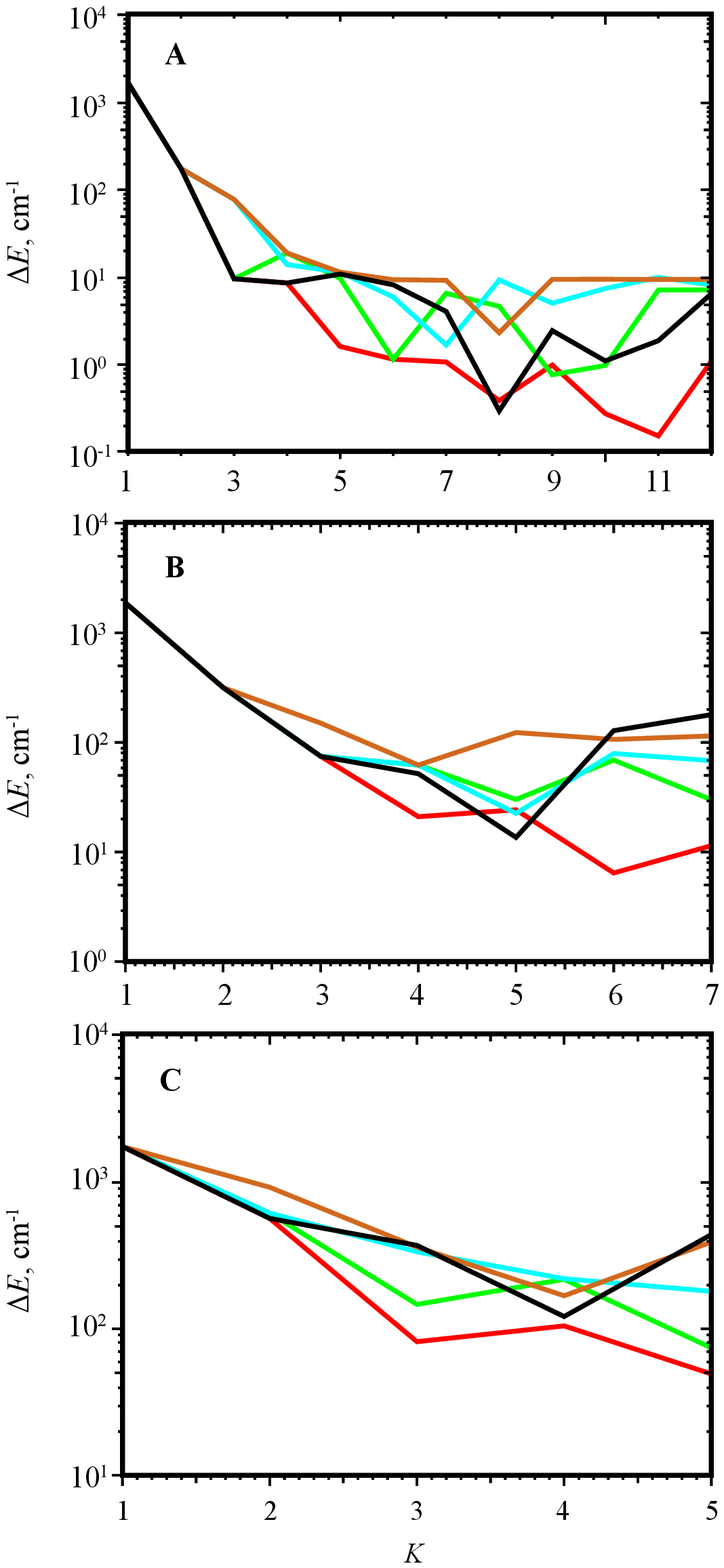}
\caption{
\begin{small} {\textbf{Simulation of quantum annealer noise.} (\textbf{A}) 1D harmonic oscillator, (\textbf{B}) O$_2$ and (\textbf{C}) O$_3$. In each panel, the classical QUBO solution error (relative to LAPACK) is plotted as a function of the number of qubits $K$ for different noise models. In general, the error for the quantum annealer (black curves) is larger than for the classical solver without noise (red curves). The quantum annealer behavior can be qualitatively reproduced by adding noise to the  classical calculation. For the 1D harmonic oscillator and O$_2$ the noise was added with scaling factors of 1 (green), 3 (blue) and 5 (brown). For O$_3$ the noise scaling factors are 3 (green), 5 (blue) and 7 (brown).}
\end{small}
}
\label{fig06}
\end{figure}

The results obtained on the quantum annealer are much less accurate than those obtained using 
the classical QUBO solver (see Figures \ref{fig01} and \ref{fig02}). We believe that this 
discrepancy is partially due to the error with which the QUBO problem is programmed in the 
hardware. The reported integrated control errors (ICE) for the hardware we used are quite 
large \cite{dw2000Q}. For example, the diagonal elements are programmed with a mean error 
0.7\% of the maximum matrix element. In the O$_2$ calculation, the maximum matrix element is 
$11.5 \times 10^3$ cm$^{-1}$ which translates into an ICE of 80 cm$^{-1}$. Moreover, the 
error has a quite broad distribution, its standard deviation is 0.8\%. This can potentially 
double the ICE, bringing it to 160 cm$^{-1}$. Upon consideration of this source of error, 
the large difference between the quantum annealer and classical results in Figure \ref{fig02} 
is no longer that surprising. To quantify the effects of the hardware errors (noise), we 
performed calculations with a classical solver where random noise was manually added to QUBO 
(see Materials and Methods). The errors (relative to LAPACK) of the classical QUBO solutions 
with different magnitudes of noise are shown in Figure \ref{fig06}. On average, introduction 
of noise increases the error. For the 1D harmonic oscillator, the default noise (using the 
reported ICE values discussed above) mimics the quantum annealer behavior. For oxygen, the 
quantum annealer behavior is well characterized if the noise is increased by a factor of three. 
For ozone, a factor of 5 or 7 is enough. As the problem size increases, larger noise scaling is 
required (to mimic the hardware performance), which implies that there could be some other 
source of discrepancy between the hardware and software solvers. Also, introduction of noise 
required an increase in the strength of the normalization weight $\lambda$ which is reflected 
in Table S1.

\section*{Discussion}

There are several places in the method where some improvements could be done and which are 
definitely worth listing. First, the functional form we used in this study is not the only 
one. For example, the initial biquadratic functional $I$, that we simplified earlier, can be 
converted to a quadratic QUBO form using multiple additional constrains. The drawback of this 
approach is that it will require additional penalty factors and ultimately will make the 
problem much harder to manage and solve. 

Second, the method would significantly benefit if the normalization condition could be 
integrated into the functional, rather than added to it. For example, one can notice that 
solution \textbf{\textit{a}} represents a point on the unit hypersphere. Thus, its position 
can be described using spherical coordinates and the angles could be approximated with qubits 
$\mathbf{q}$. However, the presence of products of sines and cosines in this kind of mapping 
results in a polynomial of high degree in $q$. This would require multiple constraints and 
associated penalty factors to convert the problem into quadratic form. The problem becomes 
difficult again.

Third, the actual scaling of the algorithm depends on the solver and its parameters. For the 
classical QUBO solver used in this work, the stopping criterion is specified by the number of 
repeats without any improvement. For the quantum annealer, the user specifies the number of 
annealing cycles or reads (see Materials and Methods). The scaling we discussed earlier is for 
the algorithm itself which excludes internal scaling effects of a particular QUBO solver.

Fourth, the physical or working Chimera graph of the quantum annealer is not perfect. The 
yield of the working graph, the percentage of working qubits (and couplers) that are present, 
is 99\% (98\%) \cite{dw2000Q}. To make it a full-yield graph, an additional software 
post-processing is performed before results are sent back to the user. The user may opt out 
of this fixing procedure, however, that reduces portability of the algorithm, since every 
machine has its own working graph. Still removing this layer might be worth exploring.

Fifth, the annealing time $t_{ann}$ is another parameter that can potentially play some role in 
the annealing process. We found that increasing $t_{ann}$ does not change results significantly. 
What affects the results more was the number of reads $N_{reads}$. Because there is a 
limitation $t_{ann} \cdot N_{reads} < 3$ seconds in the API, we chose the maximum number of 
reads per job submission: $N_{reads} = 10^4$ and $t_{ann} = 299$ $\mu$s. The maximum allowed 
$t_{ann}$ is 2000 $\mu$s but it would be good to explore larger annealing times if possible.

Sixth, it is not quite clear how to properly include the ICEs in the noise simulation tests. 
The errors are reported solely for the maximum and minimum elements of QUBO matrix and they 
are a function of annealing time. In addition, they were reported for 70\% of the annealing 
process (i.e., there is no error data at the end of annealing) \cite{dw2000Q}. For the noise 
tests reported in this work, we used the reported errors at 70\% and manually scaled them.

Seventh, the annealer temperature could be another source of error. In a recent paper \cite{temp} 
it was argued that the annealer temperatures must be appropriately scaled down with problem size 
(at least in a logarithmic way or better yet as a power law). In fact, during our study we had 
to switch from the device with 1024 qubits (DW2X) to 2048 qubits (DW2000Q). The temperature 
lowered from 15.7 ${\rm mK}$ to 14.5 ${\rm mK}$, which is almost logarithmic (it should be 
14.3 ${\rm mK}$). However, we had to perform 50 times more reads on the larger machine to 
reproduce trivial solutions for small $\lambda$ and large $c$.

Finally, the landscape of the hypothetical configuration space is defined by the problem. It 
could be that our problems have high and thick barriers on the landscape, which effectively 
disables the tunneling mechanism in the annealer and restricts exploration. Another possibility 
is that the landscape has a large number of wells and the solver jumps from one to another. 
Possibly, a very long annealing time could help to determine if this is the case.

In summary, we developed a hybrid algorithm (QAE) for the calculation of the vibrational 
spectrum of a molecule on a quantum annealer. The eigenvalue problem is mapped to the QUBO 
(Ising) problem by discretization of the expansion coefficients using qubits. The method is 
hybrid due to the scanning in a penalty (or weight) to impose wave function normalization. 
Running on the actual quantum annealer requires the additional scanning in chain penalty.
The method was applied to the ground and first excited vibrational states of two chemically 
important species: O$_2$ (oxygen) and O$_3$ (ozone). The QAE calculations based on the 
classical QUBO solver outperform those on the quantum annealer (D-Wave machine) in both 
accuracy and computational time (i.e., no supremacy of the latter one). Our tests show that 
this is partially due to the errors or noise present in the hardware. Hopefully, in the 
future it will be possible to build larger, more accurate and fully-connected quantum 
annealers. As a final note, the QAE algorithm is universal and can be used in any field 
of science or engineering to solve the real symmetric eigenvalue problem.

\section*{Materials and Methods}

We used qbsolv \cite{qbsolv} as the software classical QUBO solver and the D-Wave 2000Q 
\cite{dw2000Q} as the hardware quantum solver. The underlying qbsolv algorithm is a combination 
of Tabu search and a backbone-based method inspired by Glover \textit{et al.} \cite{glover}. The 
latter one is used for partitioning the original (large) QUBO into smaller sub-QUBOs. The only 
modification we did was to increase the span of partitioning to 1 (the hard-coded value is 
0.214). The number of repeats in the stopping criterion is $N_{rep} = 10^4$. For the excited 
states calculations we used $S_0 = 9000$ cm$^{-1}$, however 3000 and 6000 also worked well.

The hardware was accessed using qOp stack: qbsolv, DW library and SAPI \cite{qop}. Although 
qbsolv allows running sub-QUBOs on the hardware, we did not follow this approach, because the 
contribution of the D-Wave machine to the solution would be hard to estimate. Furthermore, 
qbsolv does implicit restarts and uses a classical Tabu search to refine solutions. Thus, a 
hardware calculation using the default qbsolv is actually hybrid and not fully quantum 
(for problems that fit one sub-QUBO qbsolv is completely classical). In order to bring the 
actual D-Wave performance to the surface, we removed partitioning, restarts and refinement 
from qbsolv, so that it only serves as an interface to the hardware. In addition, we raised 
the number of reads to $10^5$ (the hard-coded value is 25). Figure \ref{fig05} was prepared 
with $5 \times 10^5$ reads (half a million) and the chain penalty was 15000 cm$^{-1}$ 
(the hard-coded value is 15). The number of physical qubits on the DW2000Q is 2028 qubits 
(99\% yield) and 5903 couplers (98\% yield). However, embedding a fully connected graph leaves 
us with just 64 logical qubits.

Our code generates input QUBO matrices for qbsolv. It is written in Fortran and uses LAPACK 
\cite{lapack} as the classical numerical eigensolver (for benchmarking the QUBO results). 
Convergence studies were done for $N_{rep}$, $N_{reads}$ and $N_\lambda$ and are reported in 
Figures S2-S4 of Supplementary Materials. We did not see a strong dependence on Tabu memory, 
so we used the default values. In addition, we tested the code on $d=1$ to 5 dimensional 
harmonic oscillators and the convergence for $d=1$ to 3 in terms of number of qubits $K$ is 
given in Figure S5.

\clearpage

\clearpage
\section*{Acknowledgements}

\textbf{Funding}: A.T. and B.K.K. acknowledge that this work was done under the auspices of 
the US Department of Energy under Project No. 20170221ER of the Laboratory Directed Research 
and Development Program at Los Alamos National Laboratory. Los Alamos National Laboratory is 
operated by Los Alamos National Security, LLC, for the National Security Administration of 
the US Department of Energy under contract DE-AC52-06NA25396. We thank D-Wave Systems Inc. 
for providing access to the DW2X device at LANL and the DW2000Q device in Burnaby, Canada. 
A.T. thanks LANL for sponsoring his summer visit in 2017. D.B. acknowledges that this material 
is based upon work supported by the National Science Foundation under Grant No. AGS-1252486. 
\textbf{Author contributions:} A.T. developed the algorithm, performed the numerical 
calculations, analysis and writing of the manuscript. B.K.K. contributed to the development 
of the algorithm, analysis and writing of the manuscript. D.B. contributed to the analysis 
and writing of the manuscript. \textbf{Competing Interests:} The authors declare that they 
have no competing interests. \textbf{Data and materials availability:} All data needed to 
evaluate the conclusions in the paper are present in the paper and/or the Supplementary 
Materials. Additional data related to this paper may be requested from the authors.

\clearpage
\section*{Figures}

\makeatletter
\renewcommand{\fnum@figure}{\textbf{Fig.~\thefigure}}
\makeatother

\section*{References}

\begin{thebibliography}{9}
\makeatletter
\renewcommand\@biblabel[1]{#1.}
\makeatother

\bibitem{qcreview} Ladd, T. D., Jelezko, F., Laflamme, R., Nakamura, Y., Monroe, C. and O’Brien, J. L. Quantum computers. \textit{Nature}, \textbf{464}, 45$-$53 (2010).

\bibitem{dwave} Johnson, M. W., Amin, M. H., Gildert, S., Lanting, T., Hamze, F., Dickson, N., Harris, R., Berkley, A. J., Johansson, J., Bunyk, P. \& Chapple, E. M. Quantum annealing with manufactured spins. \textit{Nature}, \textbf{473}, 194$-$198(2011).

\bibitem{stancil} Geller, M. R., Martinis, J. M., Sornborger, A. T., Stancil, P. C., Pritchett, E. J., You, H. \& Galiautdinov A. Universal quantum simulation with prethreshold superconducting qubits: Single-excitation subspace method. \textit{Phys. Rev. A}, \textbf{91}, 062309 (2015).

\bibitem{aguzik} Aspuru-Guzik, A., Dutoi, A. D., Love, P. J. \& Head-Gordon, M. Simulated Quantum Computation of Molecular Energies. \textit{Science}, \textbf{309}, 1704$-$1707 (2005).

\bibitem{qsimreview} Buluta, I. \& Nori, F. Quantum Simulators. \textit{Science}, \textbf{326}, 108$-$111 (2009).

\bibitem{aqc} Albash, T. \& Lidar, D. A. Adiabatic quantum computation. \textit{Rev. Mod. Phys.}, \textbf{90}, 015002 (2018).

\bibitem{equiv} Mizel, A., Lidar, D. A. \& Mitchell, M. Simple Proof of Equivalence between Adiabatic Quantum Computation and the Circuit Model. \textit{Phys. Rev. Lett.}, \textbf{99}, 070502 (2007).

\bibitem{dw2000Q} D-Wave Systems Inc. QPU Properties: D-Wave 2000Q Online System (DW{\_}2000Q{\_}3). \textit{User Manual}, 09-1180A-A (2018).

\bibitem{hions1} Lin Z. \& McCoy A. B. Signatures of large-amplitude vibrations in the spectra of H5+ and D5+. \textit{J. Phys. Chem. Lett.}, \textbf{3}, 3690$-$3696 (2012).

\bibitem{hions2} Lin Z. \& McCoy A. B. Investigation of the structure and spectroscopy of H5+ using diffusion Monte Carlo. \textit{J. Phys. Chem. A}, \textbf{117}, 11725$-$11736 (2013).

\bibitem{hions3} Qu, C., Prosmiti, R. \& Bowman, J. M. MULTIMODE calculations of the infrared spectra of H+7 and D+7 using ab initio potential energy and dipole moment surfaces. In: Wilson A., Peterson K., Woon D. (eds) Thom H. Dunning, Jr. Highlights in Theoretical Chemistry, \textbf{10}, 141$-$147 (Springer, Berlin, Heidelberg, 2015)

\bibitem{hions4} Qu C. \& Bowman J. M. Diffusion Monte Carlo calculations of zero-point structures of partially deuterated isotopologues of H7+. \textit{J. Phys. Chem. B}, \textbf{118}, 8221$-$8226 (2014)


\bibitem{ch5-1} McCoy, A. B., Braams, B. J., Brown, A., Huang, X., Jin, Z., \& Bowman, J. M. Ab initio diffusion Monte Carlo calculations of the quantum behavior of CH5+ in full dimensionality. \textit{J. Phys. Chem. A}, \textbf{108}, 4991$-$4994 (2004).

\bibitem{ch5-2} Huang, X., Johnson, L. M., Bowman, J. M., \& McCoy, A. B. Deuteration effects on the structure and infrared spectrum of CH5+. \textit{J. Am. Chem. Soc.}, \textbf{128}, 3478$-$3479 (2006).


\bibitem{h3o} Petit, A. S. \& McCoy, A. B. Diffusion Monte Carlo approaches for evaluating rotationally excited states of symmetric top molecules: application to H3O+ and D3O+. \textit{J. Phys. Chem. A}, \textbf{113}, 12706$-$12714 (2009).

\bibitem{h5o2} Guasco, T. L., Johnson, M. A. \& McCoy, A. B. Unraveling anharmonic effects in the vibrational predissociation spectra of H5O2+ and its deuterated analogues. \textit{J. Phys. Chem. A}, \textbf{115}, 5847$-$5858 (2011).

\bibitem{h2clust1} Rama Krishna, M. V. \& Whaley, K. B. Structure of small molecular hydrogen clusters. \textit{Z. Phys. D}, \textbf{20}, 223$-$226 (1991).

\bibitem{h2clust2} Warnecke, S., Sevryuk, M. B., Ceperley, D. M., Toennies, J. P., Guardiola, R. \& Navarro, J. The structure of para-hydrogen clusters. \textit{Eur. Phys. J. D}, \textbf{56}, 353$-$358 (2010).

\bibitem{h2clust3} Navarro, J. \& Guardiola, R. Thermal effects on small para‐hydrogen clusters. \textit{Int. J. Quantum Chem.}, \textbf{111}, 463$-$471 (2011).


\bibitem{d2clust1} Scharf, D., Martyna, G. J. \& Klein, M. L. Isotope effect on the melting of para-hydrogen and ortho-deuterium clusters. \textit{Chem. Phys. Lett.}, \textbf{197}, 231$-$235 (1992).

\bibitem{d2clust2} Cuervo, J. E. \& Roy, P.-N. On the solid- and liquidlike nature of quantum clusters in their ground state. \textit{J. Chem. Phys.}, \textbf{128}, 224509 (2008).

\bibitem{hbs1} Curotto, E. \& Mella, M. Quantum Monte Carlo simulations of selected ammonia clusters (n=2-5): Isotope effects on the ground state of typical hydrogen bonded systems. \textit{J. Chem. Phys.}, \textbf{133}, 214301 (2010).

\bibitem{lj1} Noya, E. G. \& Doyea, J. P. K. Structural transitions in the 309-atom magic number Lennard-Jones cluster. \textit{J. Chem. Phys.}, \textbf{124}, 104503 (2006).

\bibitem{lj2} Deckman, J., Frantsuzov, P. A., \& Mandelshtam, V.A. Quantum transitions in Lennard-Jones clusters. \textit{Phys. Rev. E}, \textbf{77}, 052102 (2008).

\bibitem{vqe1} Peruzzo, A., McClean, J., Shadbolt, P., Yung, M.-H., Zhou, X.-Q., Love, P. J., Aspuru-Guzik, A., \& O'Brien, J. L. A variational eigenvalue solver on a photonic quantum processor. \textit{Nat. Commun.}  \textbf{5}, 4213 (2014).

\bibitem{vqe2} McClean, J. R., Romero, J., Babbush, R. \& Aspuru-Guzik, A. The theory of variational hybrid quantum-classical algorithms. \textit{New J. Phys.}, \textbf{18}, 023023 (2016).

\bibitem{vqe3} O'Malley, P. J. J. , Babbush, R., Kivlichan, I. D., Romero, J., McClean, J. R., Barends, R., Kelly, J., Roushan, P., Tranter, A., Ding, N., Campbell, B., Chen, Y., Chen, Z., Chiaro, B., Dunsworth, A., Fowler, A. G., Jeffrey, E., Lucero, E., Megrant, A., Mutus, J. Y., Neeley, M., Neill, C., Quintana, C., Sank, D., Vainsencher, A., Wenner, J., White, T. C., Coveney, P. V., Love, P. J., Neven, H., Aspuru-Guzik, A. \& Martinis, J. M. Scalable quantum simulation of molecular energies. \textit{Phys. Rev. X}, \textbf{6}, 031007 (2016).

\bibitem{dwProg} D-Wave Systems Inc. Programming with QUBOs. \textit{Tech. Report}, Release 2.4, 09-1002A-C (2017).

\bibitem{sdt1} Ba\u ci\'c, Z. \& Light, J. C. Highly excited vibrational levels of ‘‘floppy’’ triatomic molecules: A discrete variable representation$-$Distributed Gaussian basis approach. \textit{J. Chem. Phys.} \textbf{85}, 4594$-$4604 (1986).

\bibitem{sdt2} Light, C. \& Ba\u ci\'c, Z. Adiabatic approximation and nonadiabatic corrections in the discrete variable representation: Highly excited vibrational states of triatomic molecules. \textit{J. Chem. Phys.} \textbf{87}, 4008$-$4019 (1987).

\bibitem{sdt3} Ba\u ci\'c, Z. \& Light, J. C. Accurate localized and delocalized vibrational states of HCN/HNC. \textit{J. Chem. Phys.}, \textbf{86}, 3065$-$3077 (1987).

\bibitem{ozone1} Teplukhin, A. \& Babikov D. Efficient method for calculations of ro-vibrational states in triatomic molecules near dissociation threshold: Application to ozone. \textit{J. Chem. Phys.}, \textbf{145}, 114106 (2016).

\bibitem{ozonepes} Dawes, R., Lolur, P., Li, A., Jiang, B. \& Guo, H. Communication: An accurate global potential energy surface for the ground electronic state of ozone. \textit{J. Chem. Phys.}, \textbf{139}, 201103 (2013).

\bibitem{lapack} Anderson, E., Bai, Z., Bischof, C., Blackford, S., Demmel J., Dongarra, J., DuCroz, J., Greenbaum, A., Hammerling, S., McKenney, A., Sorensen, D. LAPACK Users' Guide. (SIAM: Philadelphia, 1999, 3rd ed.).

\bibitem{ozone2} Ndengu\'e, S., Dawes, R., Wang, X.-G., Carrington, T. Jr., Sun, Z. \& Guo, H. Calculated vibrational states of ozone up to dissociation. \textit{J. Chem. Phys.}, \textbf{144}, 074302 (2016).

\bibitem{ozonecoords1} Teplukhin, A. \& Babikov D. Interactive tool for visualization of adiabatic adjustment in APH coordinates for computational studies of vibrational motion and chemical reactions. \textit{Chem. Phys. Lett.}, \textbf{614}, 99$-$103 (2014).

\bibitem{ozonecoords2} Teplukhin, A. \& Babikov D. Visualization of Potential Energy Function Using an Isoenergy
Approach and 3D Prototyping. \textit{J. Chem. Educ.}, \textbf{92}, 305$-$309 (2015).

\bibitem{glover} Wang, Y., L\"u, Z., Glover, F., Hao, J.-K. A Multilevel Algorithm for Large Unconstrained Binary Quadratic Optimization. In: Beldiceanu N., Jussien N., Pinson É. (eds) Integration of AI and OR Techniques in Contraint Programming for Combinatorial Optimzation Problems. CPAIOR 2012. Lecture Notes in Computer Science, \textbf{7298}. (Springer, Berlin, Heidelberg, 2012)

\bibitem{temp} Albash, T., Martin-Mayor, V., \& Hen, I. Temperature scaling law for quantum annealing optimizers. \textit{Phys. Rev. Lett.} \textbf{119}, 110502 (2017). Code is available at https://github.com/dwavesystems/qbsolv

\bibitem{qbsolv} D-Wave Systems Inc. Partitioning Optimization Problems for Hybrid Classical/Quantum Execution. \textit{Tech. Report}, 14-1006A-A (2017).

\bibitem{qop} D-Wave Systems Inc. qOp toolset 2.5.1. Available from: https://www.dwavesys.com (accessed September 2018).

\end{thebibliography}
\end{document}


\raggedbottom

\section*{Supplementary Materials}

\begin{figure}[h]
\includegraphics[scale=0.5]{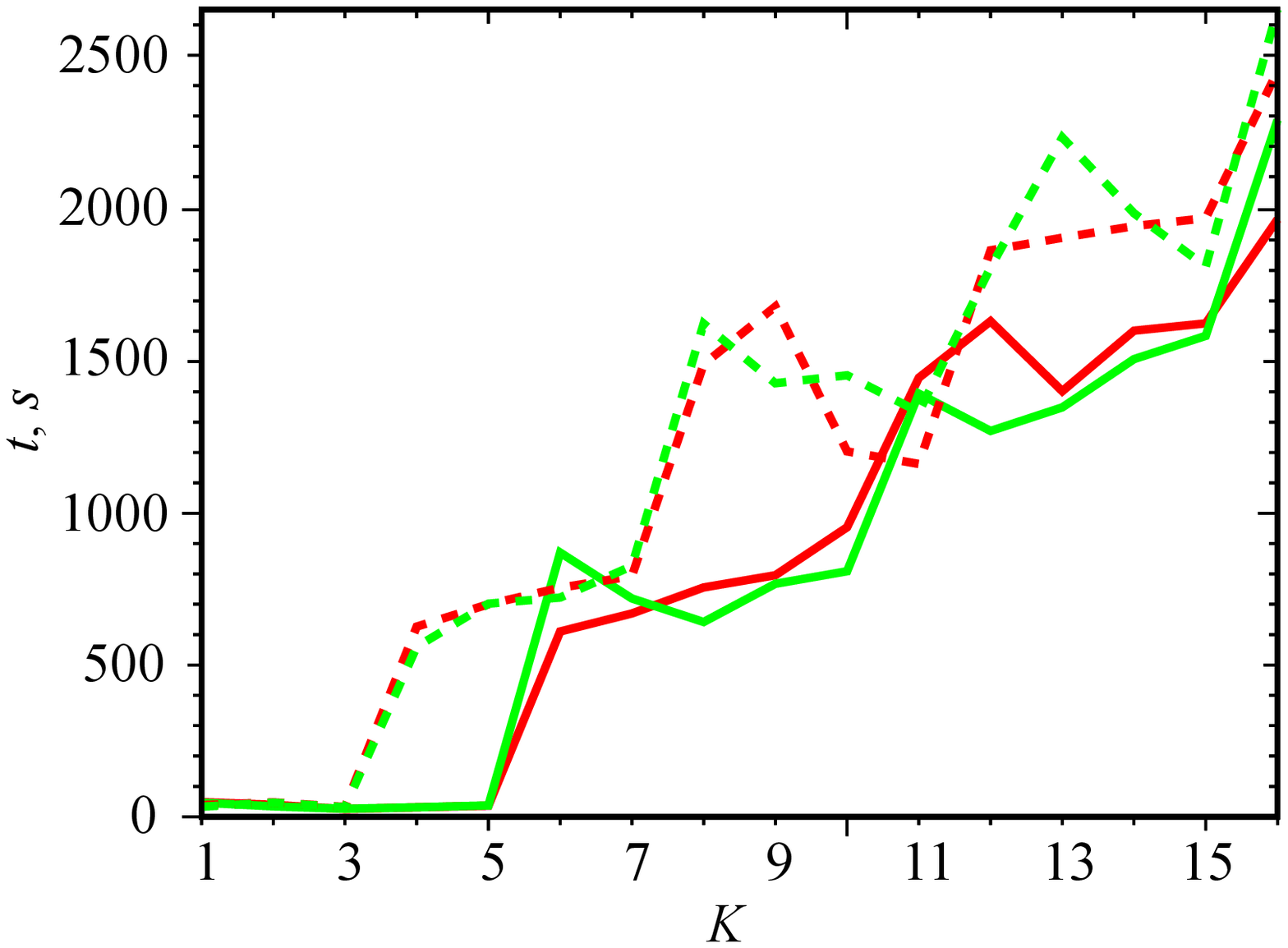}
\caption{
\begin{small} {\textbf{Computational time using a classical QUBO solver.} The molecules are O$_2$ (solid lines) and O$_3$ (dashed lines). Both ground (red curve) and first excited (green curve) states are shown.}
\end{small}
}
\label{figS1}
\end{figure}

In this document we are addressing several important questions to support the findings of 
the main text. First of all, as it was mentioned in the subsection on algorithm scaling, 
the QUBO solver may also have some additional effects on performance. Figure S\ref{figS1} 
shows a step-like runtime as a function of $K$ for the classical QUBO solver we used in this 
work (see Materials and Methods). The steps are caused by a backbone-based method inspired 
by Glover, \textit{et al.} (38) used to decompose a large QUBO instance into smaller sub-QUBOs 
and further combining these to produce a solution of the large (original) QUBO problem. The 
QUBO problems which are smaller than the sub-QUBO size (the default is 47 or is defined by 
hardware) are solved directly and are extremely fast, resulting in a negligibly small runtime 
for small $K$ in the figure. The steps become less distinguishable with increasing problem size. 

\begin{figure}[h]
\includegraphics[scale=0.5]{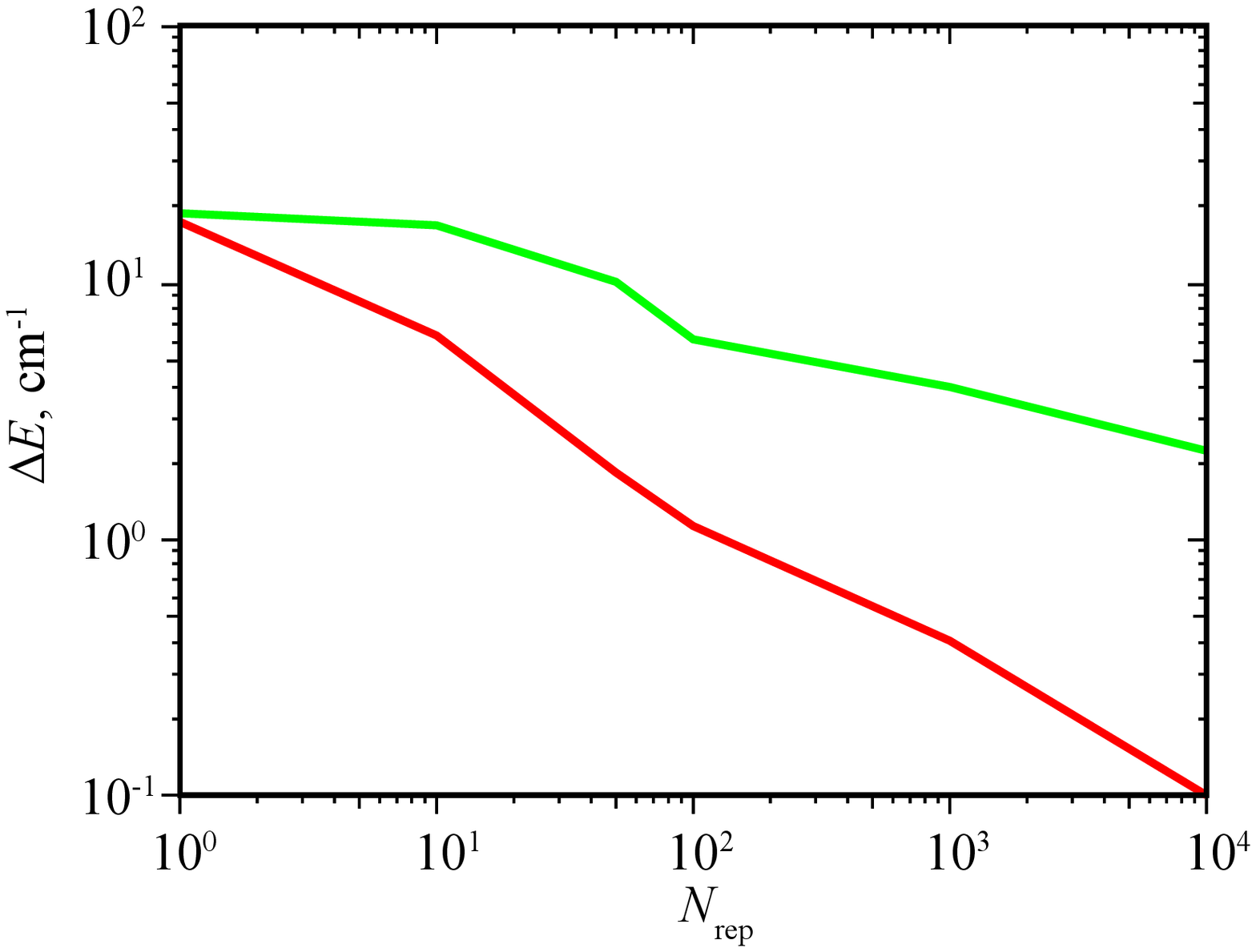}
\caption{
\begin{small} {\textbf{Convergence of ground state energies with respect to the number of repeats \textit{N}$_{rep}$ in qbsolv stopping criterion.} The molecules are O$_2$ (red curve) and O$_3$ (green curve). The classical QUBO solver was used.}
\end{small}
}
\label{figS2}
\end{figure}

Next, we report the primary results of the convergence studies for the QAE algorithm. All 
errors are computed relative to the energies of the classical numerical eigensolver LAPACK.
In Figures S\ref{figS2}-S\ref{figS4} convergence of the ground state energy is given for the
molecules O$_2$ and O$_3$, in addition to the dependence on number of qubits $K$ discussed 
earlier in the main text (see Figure 2). The basis size is the same across all 
three figures. For O$_2$, the Fourier basis size is $B = 9$ ($m_{max} = 4$). For O$_3$, the 
SDT basis size is $B = 12$ ($E_{cut} = 2000$ cm$^{-1}$). The analysis depends on the QUBO 
solver used in conjunction with QAE algorithm. 

\begin{figure}[h]
\includegraphics[scale=0.5]{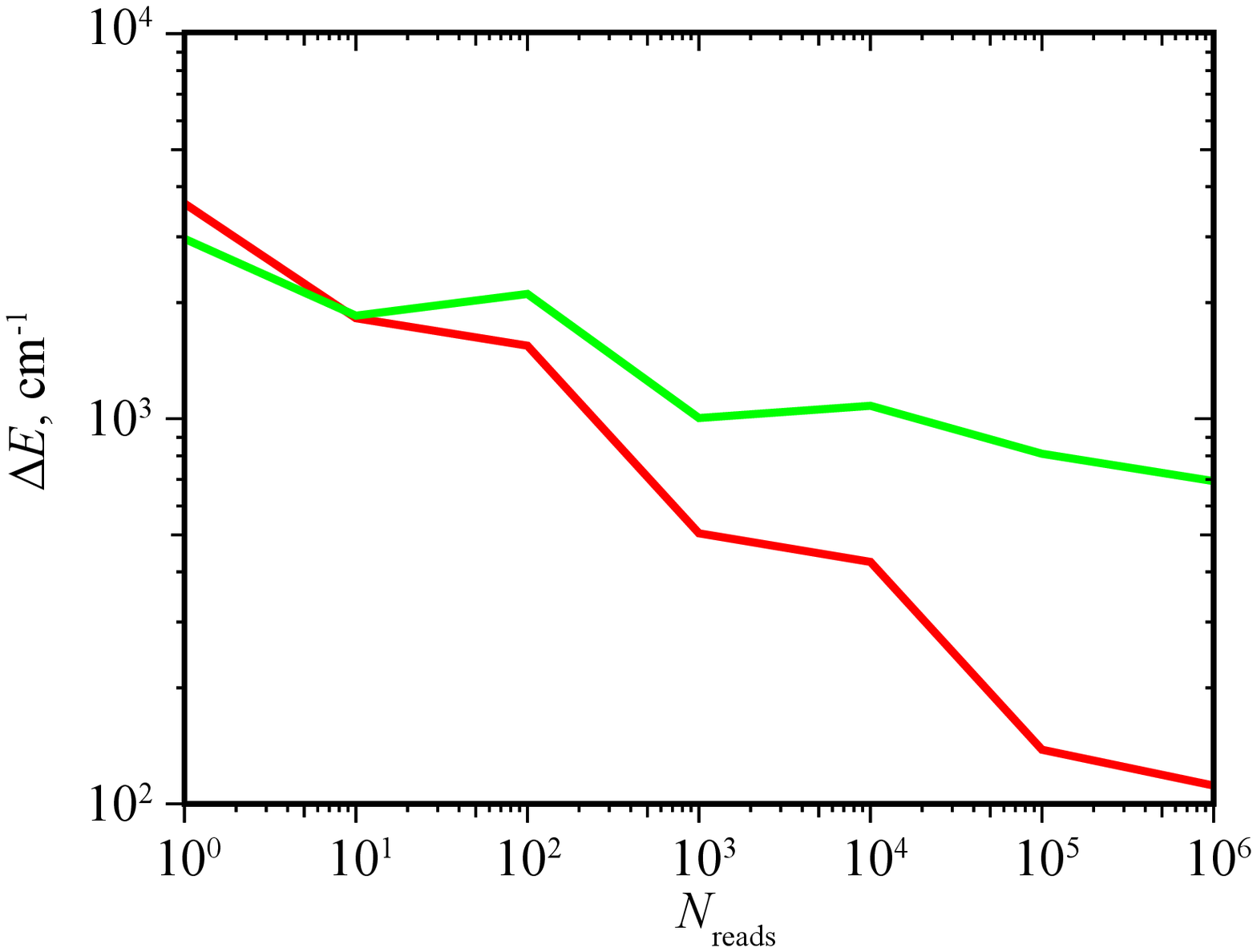}
\caption{
\begin{small} {\textbf{Convergence of ground state energies with respect to the number of reads \textit{N}$_{reads}$ in the quantum annealer.} The molecules are O$_2$ (red curve) and O$_3$ (green curve). D-Wave machine was used.}
\end{small}
}
\label{figS3}
\end{figure}

\begin{figure}[h]
\includegraphics[scale=0.5]{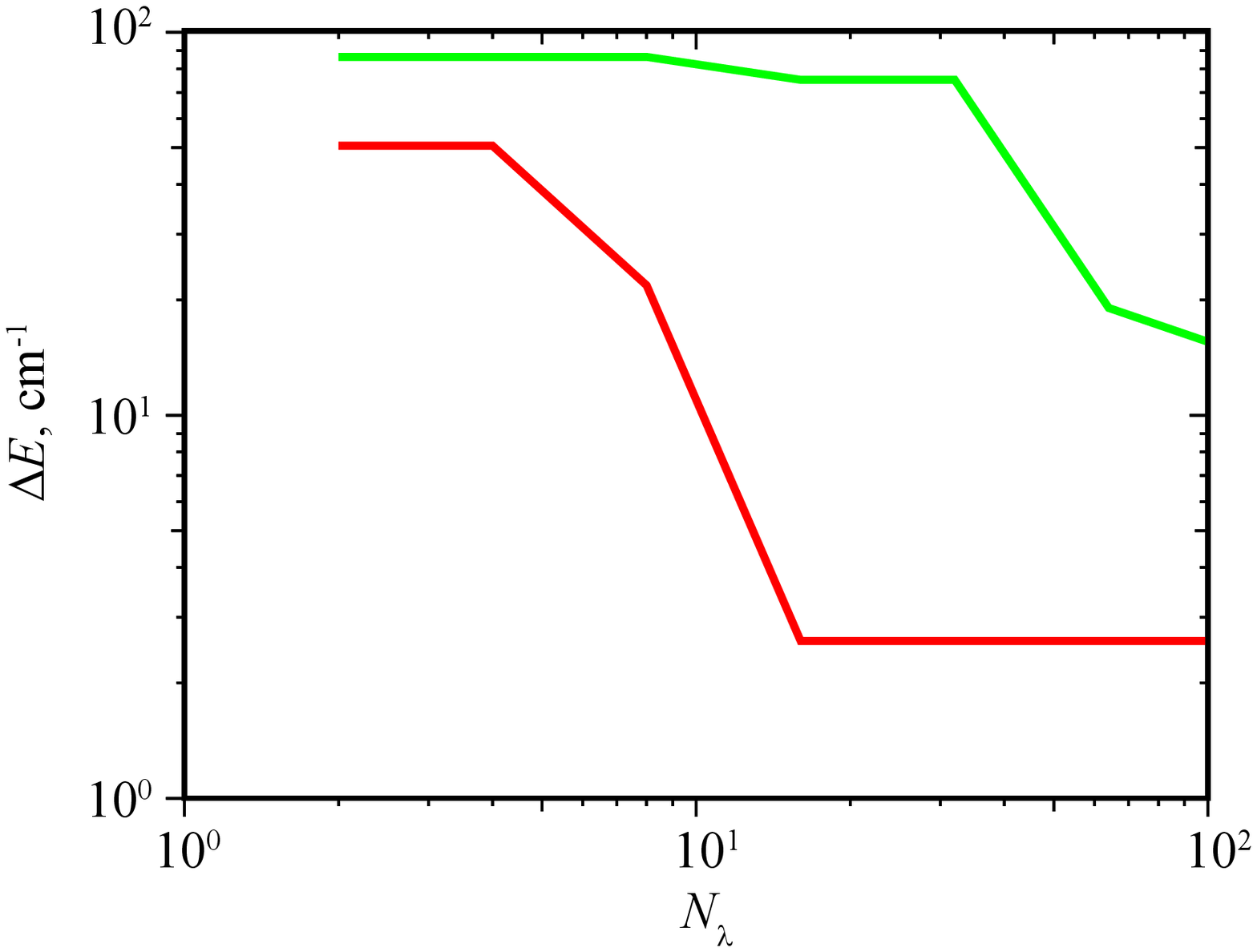}
\caption{
\begin{small} {\textbf{Convergence of ground state energies with respect to the number of steps \textit{N}$_\lambda$.} The molecules are O$_2$ (red curve) and O$_3$ (green curve). $\Delta\lambda$ was adjusted to keep the same scanning range. The classical QUBO solver was used.}
\end{small}
}
\label{figS4}
\end{figure}

For the software based classical QUBO solver qbsolv (Tabu search with subspace iteration), 
the part of qbsolv which mostly influences convergence is the stopping criterion. This 
criterion is specified by the maximum number of repeats $N_{rep}$ the solver has to perform 
without any improvement to the solution. Figure S\ref{figS2} demonstrates convergence as a 
function of $N_{rep}$. The number of qubits per coefficient is $K = 8$ for both molecules. 
For the hardware based QUBO solver (D-Wave quantum annealer), the most sensitive parameter 
is the number of reads $N_{reads}$ (number of annealing cycles). The convergence with respect 
to $N_{reads}$ is shown in Figure S\ref{figS3}. The number of qubits per coefficient is 
$K = 7$ for O$_2$ and $K = 5$ for O$_3$. The dependence on other QUBO solver parameters, 
for example, the size of Tabu memory for qbsolv or annealing time for the D-Wave annealer 
was found to be negligible.

The parameter which is solver independent and is intrinsic to the algorithm itself is the 
number of steps $N_\lambda$ for the normalization penalty $\lambda$. Figure S\ref{figS4} 
illustrates the effect of $N_\lambda$ on the energy errors. The scanning range of 2000 
cm$^{-1}$ was kept unchanged, whereas the step size $\Delta\lambda$ was decreased. Increasing 
the $\lambda$ resolution gives a better result. The initial $\lambda$ is given in Table 1. 
The number of qubits per coefficient is the same as in Figure S\ref{figS2}.

\begin{figure}[h]
\includegraphics[scale=0.8]{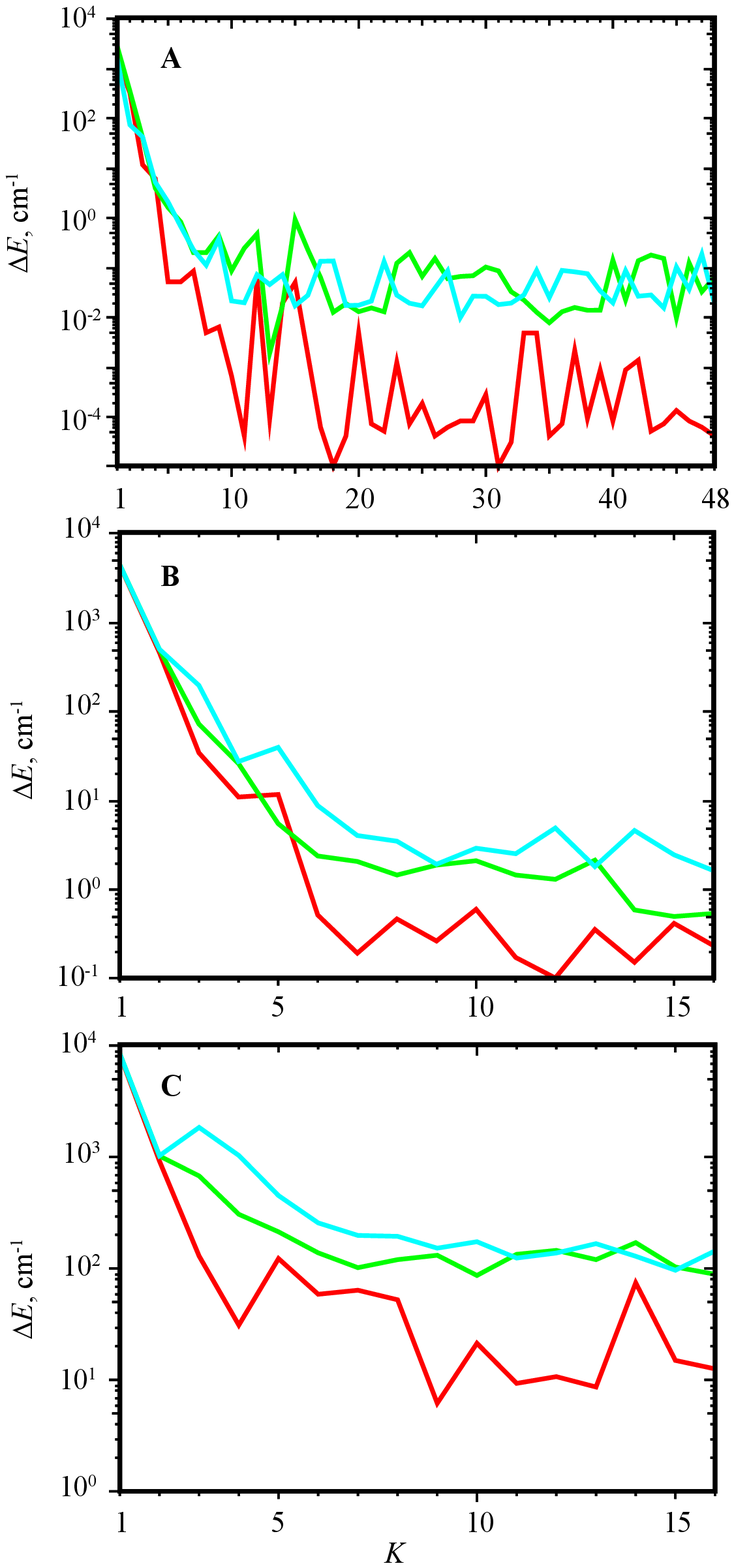}
\caption{
\begin{small} {\textbf{Ground state energy error as a function of number of qubits $K$ per expansion coefficient \textit{a}$_\alpha$.} Computed for (\textbf{A}) 1D, (\textbf{B}) 2D and (\textbf{C}) 3D harmonic oscillators. The number of cosine basis functions is $B = 3$ (red), $B = 4$ (green) and $B = 5$ (blue). The classical QUBO solver was used.}
\end{small}
}
\label{figS5}
\end{figure}

The last Figure S\ref{figS5} shows the convergence of the ground state energy error for the 
$d$-dimensional harmonic oscillator as a function of the number of qubits per expansion 
coefficient $K$. The cosine basis size is $B = 3$ ($m_{max} = 2$), $B = 4$ ($m_{max} = 3$) 
and $B = 5$ ($m_{max} = 4$). This benchmark calculation exhibits the same features observed 
in the calculations for the real systems O$_2$ and O$_3$ (see Figure 2). These 
features are an exponential drop of the error for small $K$, a plateau for large $K$ and an 
increase of the error with problem size (both in the number of dimensions $d$ and number of 
basis functions $B$).

Finally, Table \ref{table1} summarizes parameters that we used to perform 
$\lambda$ scanning for different ground state problems: initial value $\lambda_{min}$, 
number of steps $N_\lambda$ and step size $\Delta\lambda$. As indicated in the last column, 
the step size must be significantly increased for small problem sizes to avoid trivial 
solution $\Psi=0$. Parameters for first excited states of O$_2$ and O$_3$ are also given.

\clearpage
\makeatletter
\renewcommand{\fnum@figure}{\textbf{Fig.~S\thefigure}}
\makeatother

\clearpage
\makeatletter
\renewcommand{\fnum@table}{\textbf{Table ~\thetable}}
\makeatother

\begin{table}[h]
\caption{\begin{small} \textbf{Parameters for normalization penalty $\lambda$ in cm$^{-1}$} \end{small}}
\setlength{\tabcolsep}{1.0em}
\begin{tabular}{ccccc}
\hline
  
  Problem & $\lambda_{min}$ & $N_\lambda$ & $\Delta\lambda$ & $\Delta\lambda$ for small $K$ \\
  \hline
  
  O$_2$ & 780 & 10 & 10 & 400 ($K$ = 1) and 200 ($K$ = 2-4) \\
  
  O$_2$ 1st excited & 2350 & 10 & 10 & 800 ($K$ = 1) and 100 ($K$ = 2) \\

  O$_3$ & 1560 & 10 & 10 & 400 ($K$ = 1) and 100 ($K$ = 2-3) \\

  O$_3$ 1st excited & 2840 & 10 & 10 & 800 ($K$ = 1) and 100 ($K$ = 2-3) \\

  1D-harmonic & 380 & 10 & 10 & 400 ($K$ = 1) and 200 ($K$ = 2) \\

  2D-harmonic & 880 & 10 & 10 & 800 ($K$ = 1) and 100 ($K$ = 2-3) \\

  3D-harmonic & 1580 & 10 & 10 & 800 ($K$ = 1) and 100 ($K$ = 2-8) \\

  4D-harmonic & 3000 & 10 & 200 & 1600 ($K$ = 1) \\

  5D-harmonic & 3000 & 10 & 500 & 3000 ($K$ = 1) \\

  1D-harmonic + noise & 380 & 40 & 50 \\

  O$_2$ + noise & 780 & 40 & 50 \\

  O$_3$ + noise & 1560 & 40 & 50 \\

\hline
\end{tabular}
\label{table1}
\end{table}